\documentstyle[11pt]{article}

\def\bi{\begin{itemize}}
\def\ei{\end{itemize}}

\def\bc{\begin{center}}
\def\ec{\end{center}}

\def\ben{\begin{enumerate}}
\def\een{\end{enumerate}}

\def\bq{\begin{quotation}}
\def\eq{\end{quotation}}

\def\beq{\begin{equation}}
\def\eeq{\end{equation}}

\def\beqa{\begin{eqnarray}}
\def\eeqa{\end{eqnarray}}

\begin{document}

\title{Messages on Networks as a Many-Body Problem}
\author{Haralambos Marmanis\\
Division of Applied Mathematics \\
Brown University\\
Providence, RI 02912, USA}
\maketitle

\begin{abstract}
In this paper  we initiate an approach that  deals with the problem of
calculating average properties  of messages traveling on  networks, by
employing concepts   and methods that  are  used for the  study of the
many-body problem in the field of physics.  We set up a framework that
simplifies enormously the problem and,  through a concrete example, we
show how it can be applied to a broad class of networks and protocols.
\end{abstract}

\section{Introduction}
The technological advances of the  last two decades have made possible
the   construction  of   large networks    both   on   a global  scale
(e.g.  Internet) and on a  local  scale (e.g. Local  Area Networks  or
LANs, massively parallel machines).  As the  number of the constituent
nodes  and   the  number of communication    messages  increase, it is
interesting --   and probably  useful   --  to  ponder what   the {\em
effective  properties} of a given  network   will be, under a  certain
`load' of messages.  For, since  the  messages can originate  from any
node, at any time, and be sent to any other node,  it is reasonable to
expect that  one  message during its travel  from   its origin to  its
destination will be influenced by the  existence of the other messages
traveling on the same network. Hence, although a message needs a given
amount   of time to travel  between  two  arbitrary nodes  on an empty
network, it will need a  longer time to  travel the same distance on a
loaded  network. In this paper,    we  shall present a quite   general
approach that deals with issues related to this problem. The  specific
model,  presented herein,   is  an abstraction  of real  communication
networks (e.g.   optical  networks) and   real problems  on  massively
parallel  machines  (e.g.  the  {\em hot-potato}   or {\em deflection}
routing).  This generality  of  the   model naturally allows   several
extensions of itself that will be considered at a later time.

To begin  with, consider a number ($N$)  of nodes that are distributed
on  a D-dimensional surface,  that are  connected  to each  other by a
number ($N_{c}$) of links (physical or virtual), and that continuously
communicate with each   other   through these  links. The   nodes  are
enumerated  by an integer number  $i$, and  their location in physical
space  is  given      by the  coordinates    of    each   node   ${\bf
r}^i=(x_1^i,x_2^i, \ldots, x_D^i)$.  Thus, the links can be considered
as D-dimensional vectors $\vec{V}_{ij}$  whose components are given by
the usual  vector formulas, if  the coordinates  of  the two connected
nodes $i$  and   $j$ are known.    An immediate  consequence  of  this
definition is that the  vector-matrix $\vec{V}_{ij}$  is antisymmetric
(i.e. $\vec{V}_{ij}=-\vec{V}_{ji}$). The  speed of communication $c$  
(along any  of the   $i=1,2,  \ldots,D$ dimensions), among  the nodes,  
is a characteristic  of the  network and is  considered to  be known.  
Each link can carry only one message at a  time, and each node
can   send or    receive   messages   but  it    is  not  allowed   to
store-and-forward  a   message. Therefore,  the   number ($M_{t}$)  of
messages that exist  at  any time  on this network  cannot, obviously,
exceed the number of the links. We shall call the  ratio of the number
of  messages   to  the number   of    links ($M_{t}/N_{c}$) the   {\em
coefficient of saturation} or simply the {\em saturation} ($s$) of our
network.   Moreover,  each node  can send only  a number  ($M^{i}$) of
messages that is not    larger than the  number ($N^{i}_{c}$)   of its
connections with  the other nodes.  At  any instant of time, the $M_t$
messages are distributed randomly among the $N$ nodes, and each one of
them is  sent on a  different distance which is  also considered to be
random.  Each node $i$ is aware of the following three things: \ben
\item Its global location (${\bf r}^i$),
\item its connections ($\vec{V}_{ij}$), and
\item the algorithm that describes what should  be done when a message
is received. Hereafter, we will refer to this  latter algorithm as the
{\em protocol} of our network.  \een

Let a message be sent from the node-origin $i$ to the node-destination
$j$.  If  the nodes are not  connected directly, then the message will
have to reach its  destination via a number of  other nodes  that will
enable this  connection.  As far as  the operation  of this network is
concerned,  we will consider that  each message  carries the following
two things: \ben
\item The coordinates of its final node-destination ($\vec{x}^j$), and
\item its priority ($p$).  \een This information will be read at every
node    ($k$) that intervenes   during its   travel   from the initial
node-origin   to the final     node-destination.  Hence,   with   this
information and the protocol   of the network   each node is  able  to
decide which of its connections ($\vec{V}_{kl}$) will be used for each
message that arrives, and each  message will eventually arrive at  its
final node-destination  (except,    perhaps,  for  some   pathological
situations that are not of statistical significance).

It should now be clear that this network  is a good model for variants
of hot-potato  routing that can be  used by parallel machines  such as
the HEP multiprocessor (Smith, 1981), massively-parallel machines such
as the Caltech Mosaic C (Seitz, 1992), and by high-speed communication
networks (Maxemchuk, 1989). It  should also be appropriate for optical
networks (Acampora \& Shah,  1991; Szymanski, 1990; Zhang \& Acampora,
1991),  where avoiding storage of a  message is highly desirable since
optical  transmission and  switching  rates are much   faster than the
inter-conversion  between the  optical  and   electronic forms of    a
message. Despite the practical importance of hot-potato routing and of
optical networks,  the  problem of   understanding  what the   general
properties of these  systems will be and  the issue of optimizing them
is largely open.  The abstract  construction, that we described above,
may  help us to resolve some  of the problems  that arise in practice,
because it refers  to {\em continuous}   or {\em dynamic}  routing; in
contrast  to {\em batch}  routing  for which much   is known (see, for
example, Borodin \&  Hopcroft, 1985;  Kaklamanis  {\it et al.},  1993;
Borodin {\it et  al.}, 1995).  Furthermore  it  contains {\em locally}
the attribute of  greediness,  in  the sense that   the  path that  is
followed is adaptive  to the current load   of the network and  is not
predetermined.  For such a dynamic construction,  we can define states
of ``equilibrium'' where the number of transmitted messages equals, on
the average, the number of received messages. 

The new  approach is essentially  an application of  ideas and methods
that are   found in the  study of   many-body  problems. The many-body
problem is  not associated with a  specific branch of physics. Various
forms of it range from solid state to nuclear physics and from quantum
field  theory to the problem  of  turbulence. The  basic idea is  that
there are  a number of  elementary `entities'  that they interact with
each    other and that   this  interaction  is  sufficiently strong to
influence  the  dynamical evolution  of   the  system. In  our case, 
the  elementary `entities' are the  messages  that travel through  the
links of the network.  However, the  dynamics of the many-body systems
in physics is usually known and it is given  by the evolution equation
(e.g. the   Schr\"{o}ndiger  equation), whereas  in  the  many-message
system the motion  of each message is dictated  by the protocol, which
is  not an explicit  evolution  equation. Thus,  one should  attribute
probabilities to the free propagation and  the scattering process that
are based on the qualitative features of the network and the protocol.
This is not  hard for some simple  cases (see section 3), whereas  for
more  complicated cases we can always  simulate  the local dynamics of
the protocol and deduce  its scattering properties numerically. At any
rate, once these probabilities have been found, the final  result can 
be evaluated to  an  arbitrary order.

In section 2  we  will introduce  the idea  of  the propagator   for the
messages. The calculation of the latter gives us  the probability of a
message to  travel a certain distance at  a certain time. In section 3 we
perform  explicit  calculations  for a specific  network  topology and
protocol.  Finally, in section 4 we  discuss our results and suggest what
other concepts,  used in the  case of the  physical many-body problem,
could be useful in this context.

\section{Messages on Networks as a Many-Body Problem}
In this  section we will first  introduce the idea  of the propagators
for our model-network. We will consider only ``equilibrium'' states
although our formulation does allow the treatment of cases ``far from
equilibrium.''  We shall also content ourselves with the study of
spatially homogeneous networks; that is, networks for which the 
node-origin of a message does not enter the calculation explicitly.
Hence, we can always refer to messages originating at ${\bf r}=0$ and
$t=0$. Then we will introduce  the decisive role  of the  protocol in  
the dynamics of the many-message  problem. As we have already mentioned, 
if a message is going to interact with another message, it will do so 
at the  nodes. The  protocol  describes completely  this interaction, but
somewhat  differently  than the common cases  treated  in the physical
many-body problems. For in the case of, say, a many-electron system we
know the rules of the interaction in the form of  a differential equation 
whereas in the case of the many-message problem we do not have such knowledge.

\subsection{Message propagators} 
Let  $P({\bf r}_1,{\bf r}_2;t_1,t_2)$ be the {\em probability density} 
that if a message is sent from the  node at the point  ${\bf r}_{1}$ 
and at time $t_1$, then it will be received at the point ${\bf r}_{2}$ 
and at the later time $t_2$. The saturation does not appear explicitly
in this notation but we do consider $P$ as a function of the saturation 
$s$ (see below). In the  sequel, we will consider those protocols that 
are compatible with the following scattering mechanism: 
\ben
\item There is a probability $P_s(i)$ that the message will be scattered
by the node ($i$) and $1-P_s(i)$ that it  will take the  same path as in
the non-interacting case. The $P_s(i)$ turns out to be a function of the
saturation  (in some simple  cases  this quantity completely specifies
it;  see the next section),  of the coordinates,  and possibly of time
(in cases where the load is not uniform in space nor steady in time).
\item The  probability distribution  of message-paths after scattering
at $(i)$ should be independent of the message-paths before scattering;
this simply means that the message loses its `memory' of how it got to
$(i)$.  
\een

The probability that if a message is sent from node  1, at time $t_1$,
and arrives at the node 2, at  time $t_2$, can be evaluated by
summing the probabilities of all the different ways  which can be
followed  and  lead a message from (${\bf r}_1,t_1$) to (${\bf r}_2,t_2$).  
For example, the simplest path is traveling directly from ${\bf r}_1$ to  
${\bf r}_2$, without scattering from any node, and with probability
equal to one if and only if 
$(\Pi_N ({\bf r}_2 - {\bf r}_1))/c = t_2-t_1$  
and zero otherwise; hereafter $\Pi_N$ denotes the projection of the usual
Euclidean distance onto the links of the network. Another path is from  
${\bf r}_1$ to ${\bf r}_i$ ($i \neq 1,2$) with probability 
$P_o({\bf r}_{i},{\bf r}_{1},t_i-t_1)$, scattering from the node $i$ with  
probability $P_s(i)$, and finally from ${\bf r}_i$ to ${\bf r}_2$ with
probability $P_o({\bf r}_{2},{\bf r}_{i},t_2-t_i)$. Since a message loses 
its `memory' after the scattering at the node $i$, these probabilities should
be independent of  each other,  and the joint  probability  for the whole
path should be the product of  the probabilities for  each part of the
path. That is, 
\beq
\label{independent.prob}
P [ ({\bf r}_1 \rightarrow {\bf r}_i), \, 
\underbrace{({\bf r}_i,t_i)}_{\rm scattering}, \, 
({\bf r}_i \rightarrow  {\bf r}_2)  ] = 
P_o({\bf  r}_{i},{\bf r}_{1},t_i-t_1)\,
P_s(i)\, 
P_o({\bf r}_{2},{\bf r}_{i},t_2-t_i) \; .  
\eeq
Hence, the total probability $P({\bf r}_{2},{\bf r}_{1},t_2-t_1)$ will
be the sum  of the probabilities  for all the possible paths. 
However, unlike the analogous cases of its physical counterparts, in our model 
case the instant of time $t_i$ is not independent of the node location; since 
all messages propagate with the same speed. This means that only a subset of 
the possible paths is allowed to contribute for a given distance and a given 
time separation. In particular, the probability that a message will travel
from ${\bf r}_i$ to ${\bf r}_j$ is equal to zero unless 
\beq
\label{dist-time.ineq}
t_j-t_i \geq (\Pi_N ({\bf r}_j - {\bf r}_i))/c \; .
\eeq

Furthermore, and for the same reason, the discreteness of the network implies 
the discreteness of the time variable. Thus, we finally get the following series 
for the total probability 
\beq
\label{total.prob.gen}
P({\bf  r}_{2},{\bf  r}_{1}; t_2,t_1)\, =\, P_o({\bf r}_{2},{\bf r}_{1}; 
t_2,t_1)\, 
+\, \sum_{i} \sum_{n}
P_o({\bf r}_{i},{\bf r}_{1}; t_n,t_1)\, 
P_s(i)\, 
P_o({\bf r}_{2},{\bf r}_{i}; t_2,t_n)
\eeq
\[
+\; \sum_{i}\; \sum_{j}^{i \neq j} \sum_{n,m}\; 
P_o({\bf r}_{j},{\bf r}_{1}; t_n,t_1)\; 
P_s(j)\;
P_o({\bf r}_{i},{\bf r}_{j}; t_m,t_n)\; 
P_s(i)\; 
P_o({\bf r}_{2},{\bf r}_{i}; t_2,t_m)
\]
\[
+\, \underbrace{\sum}_{\rm  scattering\;  nodes}\; \; 
\underbrace{\sum}_{\rm  scattering\;  times}\;
\left\{ \cdots \right\}\; ,
\]
where the times $t_n, t_m, \ldots, \in [t_2,t_1]$ refer to the time of 
the scattering and must satisfy the above mentioned distance-time inequality  
(\ref{dist-time.ineq}).

Note that the above paragraphs show  that Feynman diagrams can be used
to accomplish the infinite  summations involved to arbitrary order. 
Nevertheless, we will not use  this `language', because it would require 
the introduction of more techniques and concepts than we really need for 
the purposes of the present paper.  Instead we will  simply invoke a 
simple notation, that is still  compact,   and write  $P_o^{ji}$  for 
$P_o({\bf  r}_{j},{\bf r}_{i}; t_j,t_i)$,  $P^{i}$ for $P_s(i)$,   and 
similarly for  the  other quantities.
For instance, equation (\ref{total.prob.gen}) will now read 
\beq
\label{simple.total.prob.gen}
P^{2   1}\, =\,  P_{o}^{2   1}\, +\,  \sum_{i}  P_{o}^{i 1}\,  P^{i}\,
P_{o}^{2  i}\, +\, \sum_{i \neq j} P_{o}^{j  1}\,  P^{j}\, P_{o}^{i j}\,
P^{i}\,   P_{o}^{2 i}\,  +\,   \cdots\;  ,     
\eeq  
where the summation over the time variable is implied by the corresponding
node.

\subsection{The Role of The Protocol}
In  the previous   sections   we have  emphasized that if a message
interacts with another  message, it will do  so  at the nodes.  The
rules of  this interaction are determined  completely by the protocol.
Thus we have here a case  quite distinct from its natural counterpart,
in the sense that we are able to specify completely the local dynamics
of `collisions.' This is a propitious feature particular to 
the many-message system and it rises the certitude that the effective 
properties of networks will be amenable not only to a calculation but 
also to an optimization (at least, in principle), by carefully designing 
an appropriate protocol or an appropriate network for a given protocol.

It turns out that, as in the case of a gas, a crucial idea is that the
joint distribution  for two messages can  be written as the product of
the two individual probabilities, i.e. it is assumed that there are no
correlations  among   the messages.  This  is   easily implemented  by
choosing  randomly the  first message  to be  sent,  from a  number of
messages that arrived simultaneously at the specific node. The Internet 
is a good  candidate for applying such a  hypothesis, but other types of
networks that have  a  significant  saturation  may also profit   from
incorporating such a principle in their protocol.

The hypothesis of random messages is not necessary though, since local
`scattering' experiments can be   performed  numerically in order to
completely  determine the nature  of  the collisions for  an arbitrary
protocol. In  particular,  this numerical simulation of  the collision
process  should perform a study of  an isolated node that is receiving
instantaneously   two   (for   a  binary  collision),    three,  four,
etc. messages.  Then the protocol will  determine the probability that
a message will  be scattered, and  consequently the calculation of the
total probability can be made.

\section{A Network Application}
What we have  presented as a general formulation, we shall now apply 
it  to  a  specific case that  is  of practical importance. 
In particular, we will give here a network application of
our model for which, under certain assumptions, we can calculate 
explicitly the propagator. To begin with, let $D=2$ and consider  
an $N \times  N$ orthogonal lattice with a uniform spacing along  
each direction. That  is, each  $x_i, y_i\;  (i=1,2, \ldots, N)$ can  
take only values  that  are integer multiples  of the grid  spacing  
$\Delta$.  The speed  of transmission for all the messages and all 
directions is   the same and equal to  $c$. 
Moreover, we let all messages have the same priority $p$.

To  complete our setting,  let the protocol  be given by the following
rules:

\begin{verbatim}

(i) Check for  messages.  
    If only one  message exists then 
       go  to step (ii) 
    else 
       choose randomly one of them and follow step (ii).

(ii)  Form the vector that joins the local node with the final
      node-destination  of the particular message.  
      If this vector is the null vector then 
         the message has arrived at its final destination.     
      Else 
         project it onto the local connection vector-matrix,  
         choose the unoccupied link with the largest projection, 
         and send the message.

\end{verbatim}

On     such a network,   we consider    the case   where  messages are
continuously sent and received in such a manner that a steady state is
approximately reached. What is the propagator for such  a system ?  In
other words, what is  the probability that   a message will reach  its
destination after a certain time has elapsed? What is the average time
that a  message needs to travel  a given distance $R$, for a specific
saturation level ($s$)  ?  What is the uncertainty or variance for this 
quantity ? Of  course, the answer to the first question encompasses the 
answers to the other two questions, but our point is that the calculation 
of such practical quantities can be carried out in detail.

\subsection{Explicit Calculation of the Message Propagator}
The starting points are the equation (\ref{simple.total.prob.gen})  
and the general characteristics of the network, as they have been 
described in  the first part of this section. That is, in order to 
evaluate $P^{2 1}$, we will evaluate the free propagator $P_o$ and  
the scattering probability $P^i$ for this network, and then we will 
substitute them in equation  (\ref{simple.total.prob.gen}).
 
We consider the form of the free propagator for this particular network 
to be given by
\beq
\label{case.free.prop}
P_o({\bf r}; t) = H(t_*)\; \sqrt{(1-s)} \left( 1. - \frac{r}{2\, N}
\right)^{s}\; \exp[- (1-s) \alpha t_{*}]\; ,
\eeq
where $t_*$ stands for $(t - \Pi_N {\bf r}/c)$, $r$ is the projection 
$(\Pi_{N}{\bf r})$, $\alpha$ is a constant, and H denotes the step Heaviside
function. The exact value of $\alpha$ can be chosen at the end of the 
calculation to be such that normalizes the total probability. In this 
heuristic formula we take into account both the geometric features of the 
network and its load (given here by the saturation $s$). 
In equation (\ref{case.free.prop}), we assume that the 
free propagator depends nonlinearly on both the saturation and the size 
of the network. Although other choices can be appropriate as well, there
is not a general method that evaluates the free propagator.
For a specific network and a specific load, which nonetheless can be
functions of space and time, the free propagator should be chosen so
that it satisfies some general criteria. In our example, for instance,
the probability peaks at the time that corresponds to an uninterrupted
travel but then falls exponentially. Moreover, as $s$ goes to zero the 
probability goes to the unit value, only for $t=\Pi_{N}{\bf r}/c$ (as it 
should), whereas $P_o$ goes to zero as $s$ goes to one.
  
The calculation of the scattering probability is based also on the general 
features of the network (loss of  memory, locality, greediness). However, 
we  have already remarked  that the current  approach  allows a local 
numerical study of the scattering process for an arbitrary  protocol, and 
if not all the conditions enforced in this example are met then the numerical
study should be preferred. Once these calculations have been made, and
the  local `dynamics'  of the system   is fully determined, we can obtain 
results regarding the global behaviour of our system.

The calculation of the scattering probability involves the calculation
of the probability that two or more messages will be at the same node,
at the same time (let us call this the event $I$), and the probability
that  more than one messages will  ask for the same  link (let us call
this the event $II$).  The random selection of a message (see the protocol 
above) turns  out to   be an  advantage  in the  current theory. It is the  
local randomness of the  network that allows  us to consider  certain events 
as independent, and  thereby to calculate the probability of the combined 
events as the product of the probabilities of each event separately.
Thus, the next step is to express explicitly the  scattering
probability in terms of the probability that a message will ask for a
certain link, and the probability that a message exists on a link. It
should be clear that, due to  the geometric symmetry and the uniform
load of the messages on the network, the former is equal to
0.25. Whereas, the latter probability is, by  definition, equal to the
saturation ($s$) of the system. Let us denote by $M_i$ ($i=1,2,3,4$)
the  event that message $i$ asks  for a  particular  link, then the
scattering probability $P^i$ should be the product of  $Pr\{I\}$ and
$Pr\{II\}$, where 
\beq
\label{bino.mess}
Pr\{I\}  = 1 - (1-s)^4 -  4 s (1-s)^3  
\eeq 
and  
\beqa Pr\{II\}  & = &
Pr\{M_1\}\, (Pr\{M_2\}\, +\, Pr\{M_3\}\,  +\, Pr\{M_4\} \\ \nonumber 
& - &   Pr\{M_2\}\,  Pr\{M_3\}\,   -\, Pr\{M_3\}\,  Pr\{M_4\}\,   -\,
Pr\{M_4\}\,  Pr\{M_2\} \\ \nonumber  
& + & Pr\{M_2\}\,  Pr\{M_3\}\, Pr\{M_4\})\; .  
\eeqa 
The first of the above equations states that the probability of having 
two or more messages is equal to unity minus the probability  of having 
no message,  minus the probability to have only one   message.  Since  
we  can  have up    to four messages  (for this particular network) and 
we know  the probability that one message will occupy a link (this is 
equivalent to  knowing the saturation)  we use the binomial  distribution 
and obtain the  final result, i.e. equation (\ref{bino.mess}). The second 
of these equations  is just a number. We have written the expression 
explicitly, in order to reveal the underlying assumption of statistical 
independence  among  the events $M_i$ ($i=1,2,3,4$). However,
$Pr\{M_1\}=Pr\{M_2\}=Pr\{M_3\}=Pr\{M_4\}=0.25$  and   consequently  we
obtain $Pr\{II\}=0.14453125$. The final conclusion is that 
\beq
\label{collision.probability}
P^i =  0.14453125\, (1 -  (1-s)^4 - 4 s  (1-s)^3)\;  , 
\eeq  
where $s$ denotes the saturation of the system.

We are now  ready to calculate   explicitly to an arbitrary  order the
propagator  as given by equation (\ref{simple.total.prob.gen}). We have 
\beqa
\label{ex.simple.total.prob}
P^{2 1}  &  =  &  P_{o}^{2 1}\,   +\, \sum_{i} P_{o}^{i   1}\, P^{i}\,
P_{o}^{2 i}\,  \\ \nonumber 
& + &  \sum_{i \neq j}  P_{o}^{j 1}\, P^{j}\, P_{o}^{i j}\, P^{i}\,  
P_{o}^{2 i}\, \\ \nonumber 
&  + & \sum_{i \neq  j \neq k} P_{o}^{k 1}\, P^{k}\, P_{o}^{j k}\, 
P^{j}\, P_{o}^{i j}\, P^{i}\, P_{o}^{2 i}\,  \\ \nonumber 
&  + & \cdots\; .  
\eeqa  
Nevertheless, equation (\ref{collision.probability}) indicates that  $P^i$  
is independent of $i$ and therefore can  be taken  outside the summation
symbols. Furthermore, we will assume that each free propagator in each sum
is equal to $P_o({\bf r}_{2 1}/n; t)$, where $n$ equals the order of the 
summation. That is, if we consider the first sum over a single node then $n=1$, 
if we consider the second sum over two nodes then $n=2$, etc. This assumption
allows us to take into account all the orders of the summation. Hence, the
summed products can be factored out and give a single sum. In fact, it is not 
hard to show that the final result is
\beq
\label{ex.result}
P^{2 1} = H(t_*)\, \exp(-\alpha (1-s) t_*) \sum_{n=1}^{N^2} (1-s)^{n/2} 
\left( \frac{P^i N}{n} \right)^{n-1}\, \left(1\, -\, \frac{r}{2 n N} 
\right)^{n\, s}\; . 
\eeq
The above sum converges extremely rapidly for given values of $s$ and $N$.
According to (\ref{ex.result}) as the saturation of the network increases
the probability that a message will travel a certain distance at a certain
time becomes a weak function of time. On the other hand, as the saturation 
becomes negligible the effect of multiple scattering vanishes.   

\section{Discussion}
We have presented a new framework for the study of problems related to the
effective properties of networks, when many messages are communicated at 
the same time through their links. Although we have given a result obtained 
somewhat heuristically for a simple problem, it should be clear that real-world
problems, of the types mentioned in the introduction, can be tackled by the
same method. The summations in (\ref{simple.total.prob.gen}) need not be 
simplified and the calculation can be done numerically up to several orders.
Numerical evaluation of these terms is to be preferred, since it allows 
greater flexibility and speed. It is our belief that analytical results (such
as the one given in section 3) are important to reveal the qualitative behaviour
of the system, however numerical investigations should be superior in accuracy;
especially for moderate sized networks. Such numerical experiments are currently
undertaken and will be presented elsewhere. These numerical experiments will 
imitate the local behaviour of the dynamics and obtain the probability of 
free propagation and the scattering probability.  

Unlike its physical counterpart, the local dynamics of the many-message system
is not universal. Messages interact with each other according to a specific 
protocol, which we are able to change at will. Thus each protocol should be 
studied separately for a specific network. The protocol need not even be 
the same for the whole network. In that case, a partition of the network
may be necessary in order to handle correctly the interactions. 

For networks in a state of equilibrium, such as the one of our example, other
methods and concepts of statistical mechanics can also be of importance in
describing their collective properties. The notion of entropy, energy, 
temperature, and other thermodynamic concepts can be applied in these cases.
For example, it may turn out that certain networks have a critical 
``temperature'' above which the performance of the network degrades considerably.

We believe that the ideas presented will be developed further and provide 
a systematic way of dealing with large networks. The ultimate goal of these
interdisciplinary study is the optimization of existing networks by altering
the protocol and the selection of the best network structure for a given 
class of protocols.

\section*{Acknowledgments}
It  is  a pleasure   for the author to  acknowledge the fruitful 
discussions with R. M. Kirby and V. Dukic.

\end{document}